\documentclass[3p,times,twocolumn]{elsarticle}
\usepackage{ecrc}
\usepackage{graphicx}
\usepackage{aas_macros}
\usepackage{float}

\volume{00}

\firstpage{1}

\journalname{Nuclear Physics B Proceedings Supplement}

\runauth{}

\jid{nuphbp}

\jnltitlelogo{Nuclear Physics B Proceedings Supplement}

\usepackage{amssymb}

 \usepackage{lineno}

\usepackage[figuresright]{rotating}

\begin{document}

\begin{frontmatter}



\dochead{}

\title{The multi-frequency multi-temporal sky}


\author{Paolo Giommi}

\address{ASI Science Data Center (ASDC), Via del Politecnico snc, I-00133 Roma, Italy}

\begin{abstract}
Contemporary astronomy benefits of very large and rapidly growing amounts of data in all bands of the electromagnetic spectrum, from long-wavelength radio waves to high energy gamma-rays. Astronomers normally specialize in data taken in one particular energy window, however the advent of  data centers world-wide and of the Virtual Observatory, which provide simple and open access to quality data in all energy bands taken at different epochs, is making multi-frequency and multi-epoch astronomy much more affordable than in the past. New tools designed to combine and analyze these data sets are being developed with the aim of visualizing observational results and extracting  information about the physical processes powering cosmic sources in ways that were not possible before.
In this contribution blazars, a type of cosmic sources that emit highly variable radiation at all frequencies,  are used as an example to describe the possibilities of this type of astronomy today, and the discovery potential for the near future.
\end{abstract}


\end{frontmatter}


\section{Introduction}
\label{intro}
Modern astronomy rests upon highly technological ground and space-based observatories that are capable of probing the sky with high sensitivity in almost all bands of the electromagnetic spectrum.  As a consequence,  extremely large and rapidly growing amounts of high-quality digital data are being accumulated. Archive data centers, that often openly provide ready-to-use data products based on consolidated data format like FITS, together with the rapid increase of computing power and network communication speed, and the existence of world-wide initiatives like the Virtual Observatory (VO)  \citep[see e.g.][]{VO} are providing unprecedented opportunities to obtain high quality multi-frequency data. 

A new era of scientific discovery, based on large amounts of archival and fresh data covering the entire electromagnetic spectrum and accumulated over a very wide time interval, has started.
 
Existing digital archives typically include astronomical data of one of the following types 

\begin{enumerate} 
\item data produced as part of diverse and unrelated scientific projects proposed by single observes to one specific astronomical facility
\item data from large surveys of the sky carried out in different energy bands
\item data from short or long-term monitoring of specific sources
\item data taken as part of large multi-observatory programs to simultaneously observe specific targets in specific energy bands.
\end{enumerate}

In this contribution I use blazars, a special type of extragalactic sources that emit highly variable radiation across the electromagnetic spectrum,  to illustrate how this new opportunity of accessing large amounts of spectral and timing data is currently exploited in terms of techniques for visualization and analysis. I also briefly describe some new software tools, developed within the VO or related activities, that can be used to efficiently retrieve and analyze multi-frequency multi-temporal archival data. 

\section{Blazars as an example of multi-frequency multi-temporal data analysis}
\label{blazars}

Blazars are a special type of Active Galactic Nuclei (AGN) that are known to be strong emitters in all bands of the electromagnetic spectrum. These peculiar sources, known since the 
discovery of AGN fifty years ago \cite{Schmidt}, display very unusual properties like superluminal motion and are the most variable persistent sources in the extragalactic sky.
The extreme properties of blazars are thought to be the result of emission from charged particles interacting with a magnetic field in a jet of plasma that moves at relativistic speeds and 
happens to point very close to the line of sight  \citep[see][ for a review]{UP95}. These are conditions that can happen only rarely, and that 
is why only about 3,000 blazars are known \citep{bzcat}, compared to over one million AGN (http://quasars.org/milliquas.htm).

\begin{figure}[h!]
\includegraphics[width=19pc]{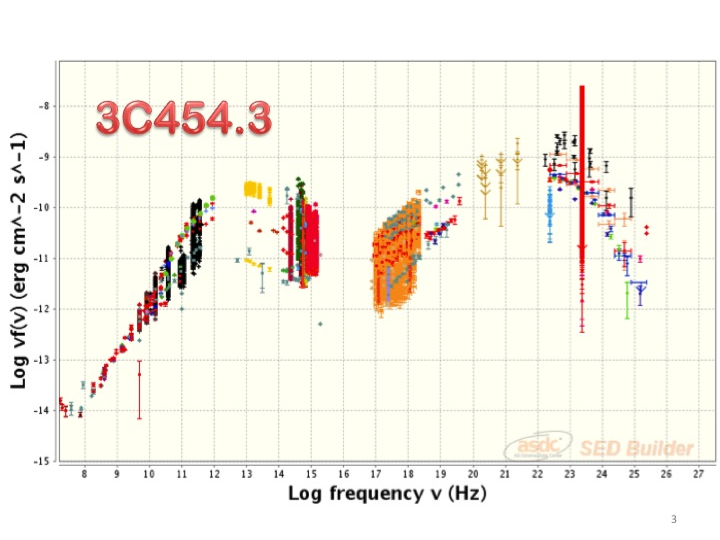}
\caption{The SED of the blazar 3C454.3 built including over 30,000 multi-frequency independent measurements. Note the extremely large variability at optical, UV, X-ray and especially at gamma-ray energies where the brightest measurement is about 10,000 larger than the weakest detection.}
 \label{SED3c454.3}
\end{figure}

Over the past several years blazars have been observed, often repeatedly, at all frequencies; in some cases, especially in the radio and optical bands, some of the brighter ones have been monitored for long periods. Consequently there are many databases and catalogs that include measurements of blazars at all frequencies (e.g. radio,  mm, IR, optical, UV, X-ray and gamma-ray). 

The broad-band emission in blazars is traditionally represented as Spectral Energy Distributions (SED), that is plots of intensity (usually flux density multiplied by frequency, $\nu f(\nu)$, (or luminosity, 
$\nu L(\nu)$) versus energy or, equivalently, frequency - $\nu$, of the emitted radiation.  As an example, figure \ref{SED3c454.3} shows the SED of the blazar 3C454.3,  currently one of the most 
densely  populated existing SED as it includes approximately 30,000 independent flux measurements collected over a time period of more than thirty years. 
A large fraction of the data shown in Fig. \ref{SED3c454.3} comes from monitoring programs and from on-line databases like UMRAO (dept.astro.lsa.umich.edu/datasets/umrao.php) at 5, 8 and 
14.5 GHz, OVRO (www.astro.caltech.edu/ovroblazars)\citep{Richards} at 15GHz, Mets\"{a}hovi (metsahovi.aalto.fi/en/research/projects/quasar/) at 37~GHz,  SMARTS (www.astro.yale.edu/smarts)\citep{Bonning} in the optical and infrared bands, WEBT (www.aoto.inaf.it/blazars/webt) \citep{Villata} at optical, IR and radio frequencies, the BeppoSAX and Swift data bases in the X-rays,  and Fermi in the gamma-ray band (www.asdc.asi.it/mmia).  The 1GeV light-curve (that appears as a vertical line at 2.4 10$^{23}$ Hz in Fig. \ref{SED3c454.3}) was built with Fermi-LAT data using the adaptive-bin method developed by \cite{Lott}

Another important example of multi-frequency data acquisition is the organization of campaigns of simultaneous observations of one or more sources involving several different facilities. The data collected in these cases are more homogeneous. An example of this approach is the Planck, Swift, Fermi and ground-based simultaneous observations of a large sample of blazars, including 175  sources selected according to four different criteria in the radio, X-ray and gamma-ray band \citep{GiommiPlanck}.

\begin{figure}[h!]
\includegraphics[width=19pc]{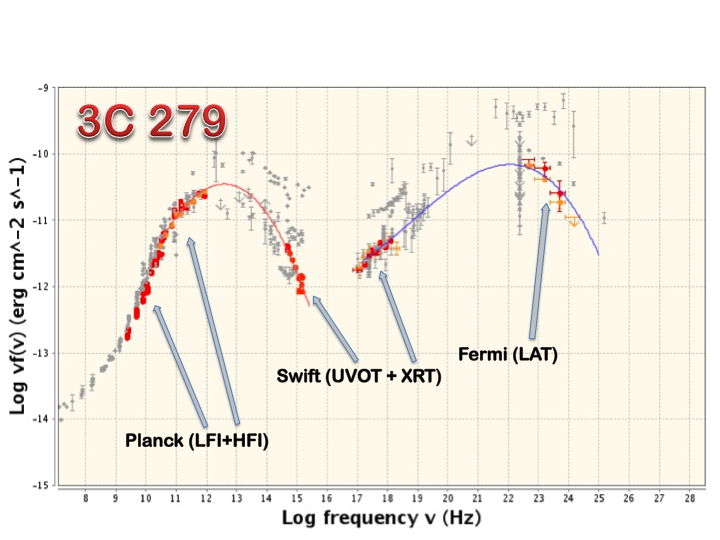}
\caption{The SED of the blazar 3C279 built with simultaneous  Planck, Swift and Fermi data \citep{GiommiPlanck}, shown as red symbols, and with non-simultaneous archival data appearing as gray points.}
 \label{SED3C279}
\end{figure}

Figure \ref{SED3C279} shows the SED of the source 3C279 taken from  \cite{GiommiPlanck} which includes {\it simultaneous data} covering a spectral range of 15 orders of magnitudes. The simultaneous measurements from the instruments of Planck (LFI, HFI) \cite{Tauber}, Swift (UVOT, XRT) \cite{Gehrels} and Fermi (LAT) \cite{Atwood}  are shown as red points, while quasi-simultaneous data (i.e. observations carried our within two months of each other) are plotted as orange points. Archival data taken at different random times appear as gray points. Simultaneous data is  clearly crucial for measuring the parameters related the emission process, like  the energy where the emitted power peaks and the intensity level of the peak.

As Figs. \ref{SED3c454.3} and \ref{SED3C279} demonstrate, the amplitude of variability in blazars is a strong function of the energy where the emission occurs, ranging from a factor of a few in the radio band, and up to a factor 10,000! at 1GeV. This dependence requires  that the time scale of 
variability in each energy band must be properly taken into consideration when dealing with 
multi-frequency data that is not simultaneous. Clearly, observation times must be visualized with methods that go beyond simple spectral distributions like that of Fig. \ref{SED3c454.3}. 

\begin{figure}[h!]
\includegraphics[width=20pc]{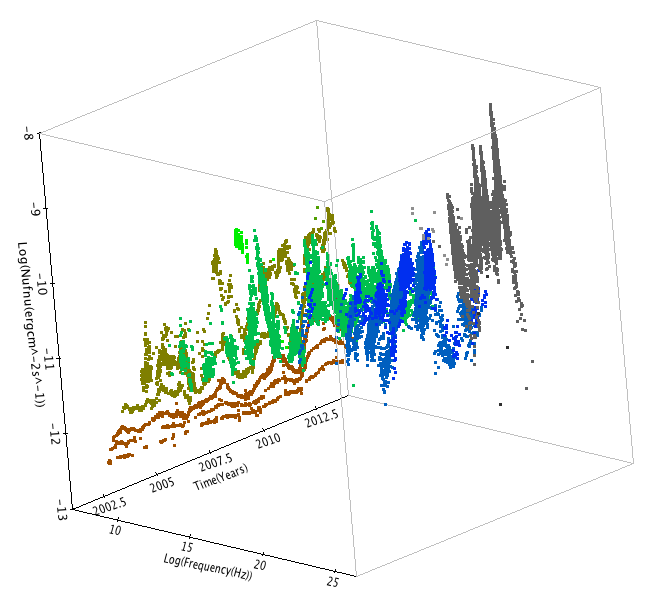}
\caption{The multi-frequency emission of the blazar 3C454.3 covering the period 2000 - 2013 is represented as a 3D plot generated using TOPCAT, a popular application developed within
Virtual Observatory projects.}
 \label{TSED3c454.3}
\end{figure}

\begin{figure}[h!]
\includegraphics[width=19pc]{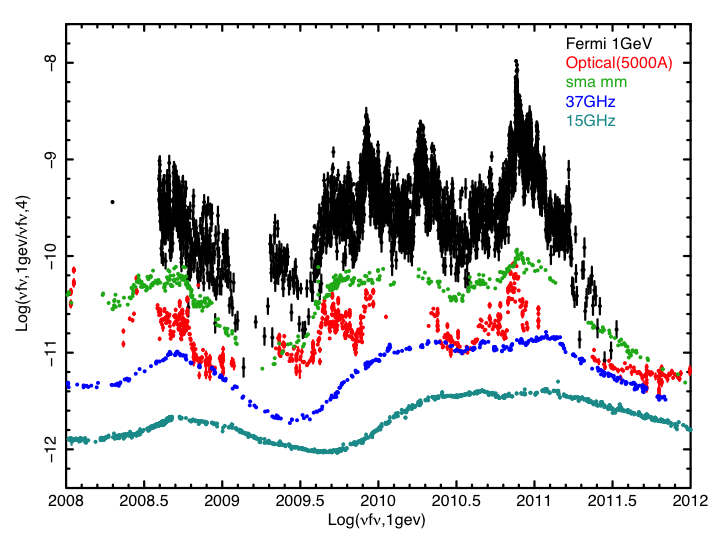}
\caption{The  time evolution of  the emission from 3C454.3 at different frequencies is shown as a 2D plot. Only a limited number of energy bands can be shown this way.}
 \label{MWlightcurves}
\end{figure}

\section{SED software tools}
\label{tools}

As described above the remarkable number of measurements now accessible and the very large intensity variations observed in several well-known sources require that the interpretation of the physical emission processes is carried out  by analyzing the SEDs in the time domain.  Until recently, however, existing software packages for building SEDs did not take into account of time. This is rapidly changing and new tools capable of handling the time variable are appearing or are planned for the near future. 

Figure \ref{TSED3c454.3} is an example of how the time dependance of the energy distribution of a source (3C454.3 in this case) can be visualized as a 3D plot, illustrating in a single picture 
how time scales, flare details, variability amplitudes and time lags vary across the electromagnetic spectrum.
This plot was generated using the TOPCAT application (http://www.star.bris.ac.uk/$\sim$mbt/topcat), which is a widely used interactive graphical viewer developed as part of some UK and Euro-VO projects.

In the following I briefly describe two new software tools that can be used to download multi-frequency measurements from many different catalogs, databases and sky surveys, and build and analyze blazar SEDs.

\subsection{The IRIS SED analysis tool}

\begin{figure}[h!]
\includegraphics[width=19pc]{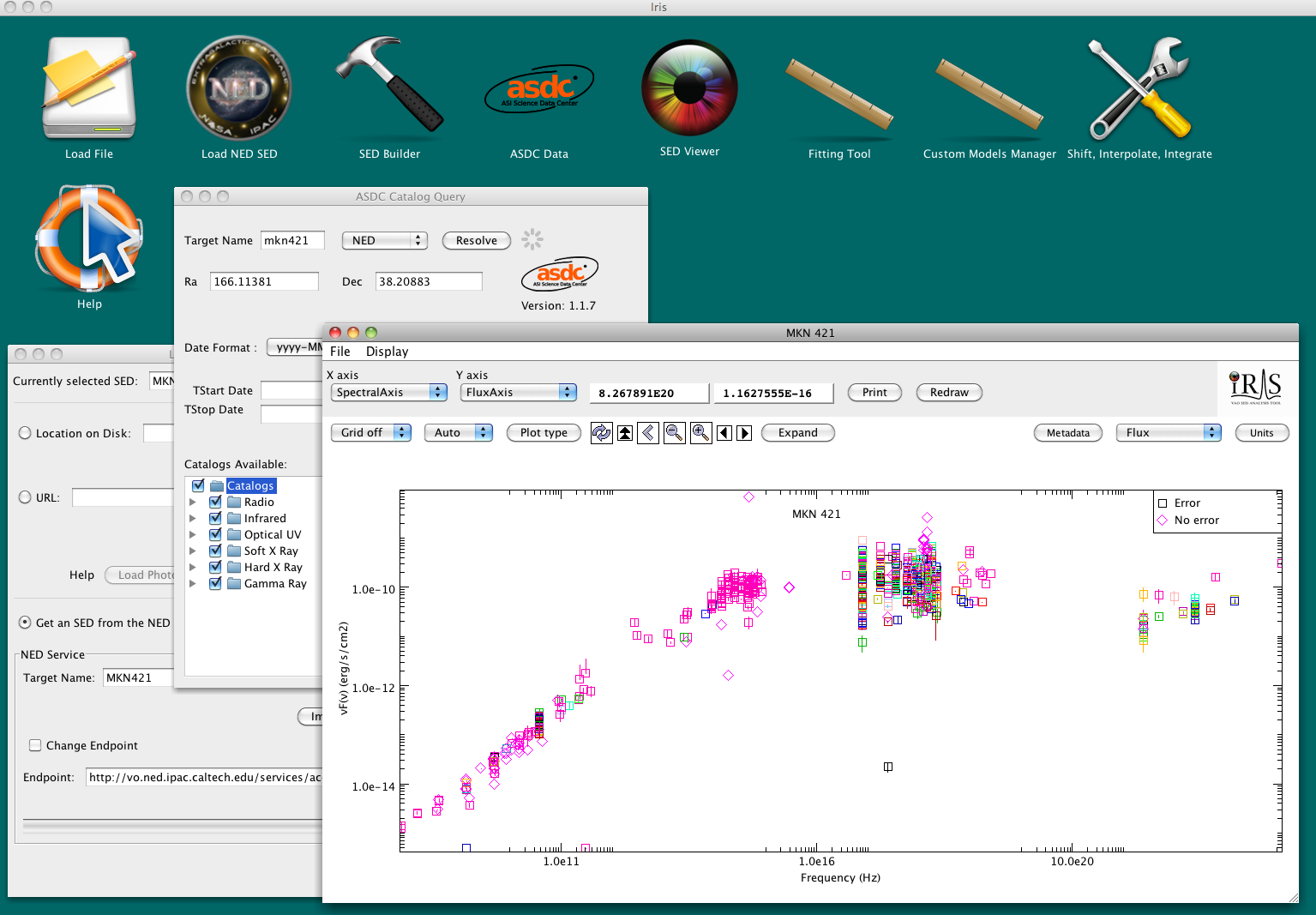}
\caption{The  IRIS application (V2.0) showing the SED of the blazar MKN421 built with flux measurements taken from NED and ASDC.}
 \label{iris}
\end{figure}

IRIS is a JAVA application developed as part of the activities of the Virtual Astronomical Observatory (VAO), the US contribution to the world-wide VO initiative.  
IRIS can retrieve data using VO protocols from the National Extragalactic Database (NED) and from the ASI Science Data Center (ASDC). It can be used to plot and fit
Spectral Energy Distributions in a number of ways.  As an example Fig. \ref{iris} shows a session 
of IRIS showing the SED of the blazar MKN421 displayed as a $\nu$f($\nu$) vs $\nu$ plot.

The current version of this desktop application (V2.0) only allows limited control of the time variable, and visualization must be done using energy or frequency (in various units) on 
the X-axis. 
The application can be downloaded from the VAO web pages at the following link http://www.usvao.org/science-tools-services/iris-sed-analysis-tool/

\subsection{The ASDC SED builder}

The ASDC SED builder is a web-based application developed at the ASDC (www.asdc.asi.it). 

\begin{figure}[h!]
\includegraphics[width=19pc]{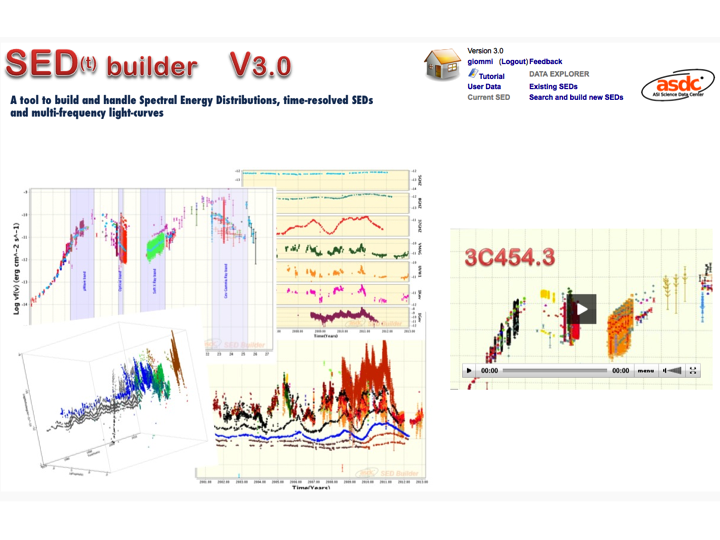}
\caption{The ASDC builder V3.0 available on the WEB at http://tools.asdc.asi.it/SED}
 \label{TEDTool}
\end{figure}
The current version (V3.0, see Fig. \ref{TEDTool}) allows users to build SEDs using data from a large number of catalogs, on-line services, also in combination with personal data. This version of the tool can handle time resolved SEDs and multi-frequency light-curves.  The service can be accessed at tools.asdc.asi.it/SED.

\subsection{SEDs and the time domain}

\begin{figure}[h!]
\includegraphics[width=19pc]{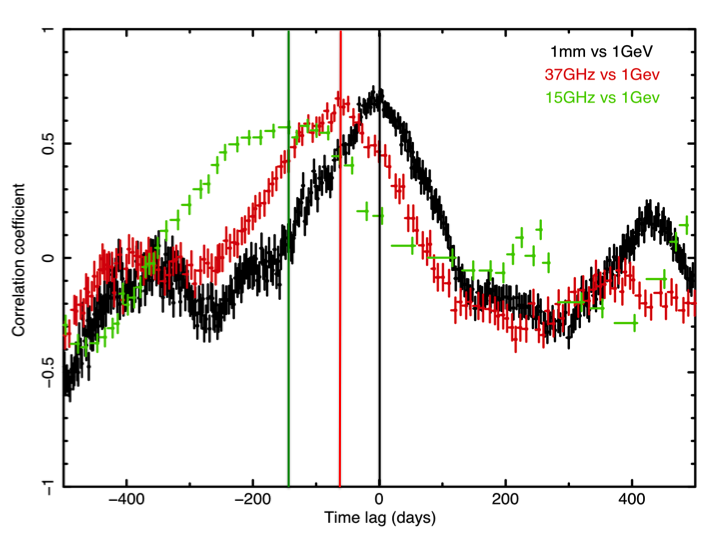}
\caption{The Z-transformed discrete cross-correlation function of the emission from 3C454.3 in different energy bands, compared to the flux emitted at 1 GeV. Significant correlation is clearly present with time lags that range from nearly zero to several weeks, depending on the frequency considered.}
      \label{zdcf3c454.3}
\end{figure}

\begin{figure}[h!]
\includegraphics[width=21pc]{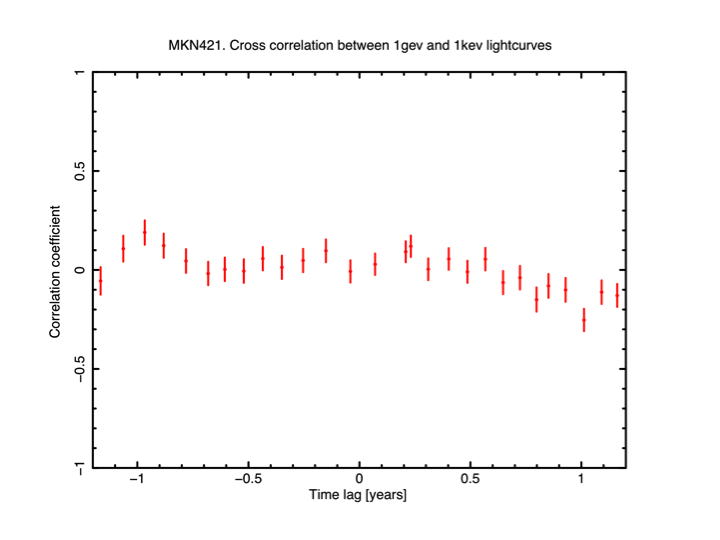}
\caption{The Z-transformed discrete cross-correlation function of the X-ray (1keV)  and Gamma-ray (1GeV) emission of the blazar MKN 421. No correlation is present between the two energy bands.}
      \label{zdcfmkn421}
\end{figure}
One important question in the analysis of the multi-frequency emission in cosmic sources is whether
the emission in different energy bands is correlated. In almost the totality of cases the measurements available are sparse and not uniformly sampled.  An efficient method of measuring the amount of correlation between the emission in two energy bands with sparse data is the   Z-transformed Discrete Correlation Function \citep[ZDCF, see e.g.][]{alexander}. 

Fig. \ref{zdcf3c454.3} shows the ZDCF of the emission form 3C454.3 at 1mm, 37GHz and 15GHz compared to the gamma-ray emission at 1GeV. The fluxes are clearly correlated but with time lags that range from approximately 0, for the case of the mm band, to several weeks depending on the frequency in the radio band. 
Note from Fig. \ref{MWlightcurves} that, although the black (1GeV)  and green (1mm) light-curves show the same overall behavior in terms of peaks and minima occurring approximately at the same time, the emission at 1GeV displays more structured variability, reflecting different details in the emission mechanism. The optical light curve (red points) follows the 1GeV light curve also in the fine detail, although data in this band is certainly more sparse than at other frequencies.  

Figure \ref{zdcfmkn421} gives the discrete correlation function of the X-ray (1~keV from Swift-XRT data) and  gamma-ray (1~GeV from Fermi-LAT data) emission from MKN421 recorded between the summer 2008 and spring 2012. No correlation is observed in this case. This is interesting since MKN 421 is  a blazar of the HBL type, that is a blazar that radiates up to the TeV band, and the radiation from this type of sources  is often interpreted as due to a single homogeneous component. The complete lack of correlation between the flux emitted in the X-ray and gamma-ray band challenges this simple interpretation.

An efficient and novel way of representing fast and large variations in the energy distribution of cosmic sources is to run in a sequence a set of frames, each representing the status of the SED in a particular time interval, like in a movie. This, of  course, is possible only if a sufficiently large number of measurements across the electromagnetic spectrum are available at all times. This requirement is already satisfied for a small number of bright 
sources today; the very rapid increase in data production that we are experiencing ensure that many others will follow in the future.

\section{Conclusions and prospects for the future}
\label{conclusion}

Significant progress in the visualization and analysis of multi-frequency multi-temporal astronomical data has been made recently, mostly as part of the VO world-wide initiative and of related activities. 
This is true both in terms of availability of data from catalogs, databases and surveys, as well as in the development of new software tools. The discovery potential offered by the exponential increase of available data, computer power and communications speed is extremely large. In this contribution I described some examples of the multi-frequency data sets and tools that are available today.  
Of course there is ample room for improvements and much more progress is expected in the near future.

\begin{figure}[h!]
\includegraphics[width=18pc]{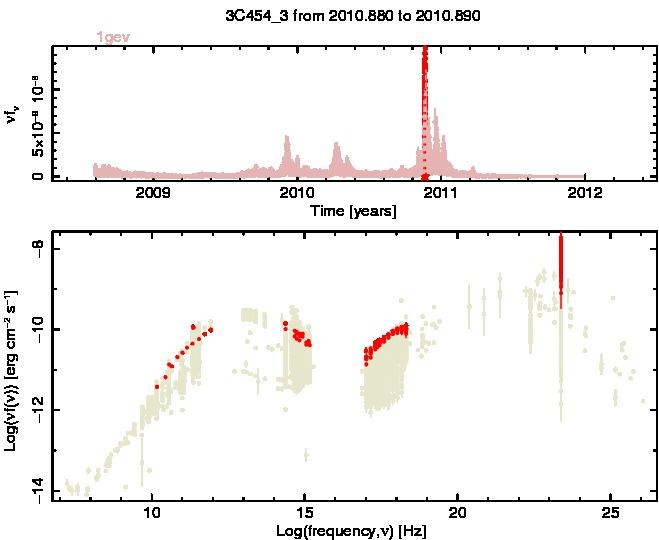}
\caption{A frame from a SED movie of the blazar 3C454.3 ceovering the period 2008-2013 built using the prototype described in the text. The top panel shows how the 1GeV flux varies as a function of time. 
The red dashed vertical lines mark the time interval considered in this frame (that is between  2010.880 and 2010.890). Spectral data taken during the same time interval, when the source showed the maximum flux in the gamma ray and in most other energy bands, are shown in the bottom panel as red symbols.}
 \label{MovieFrame1}
\end{figure}

\begin{figure}[h!]
\includegraphics[width=18pc]{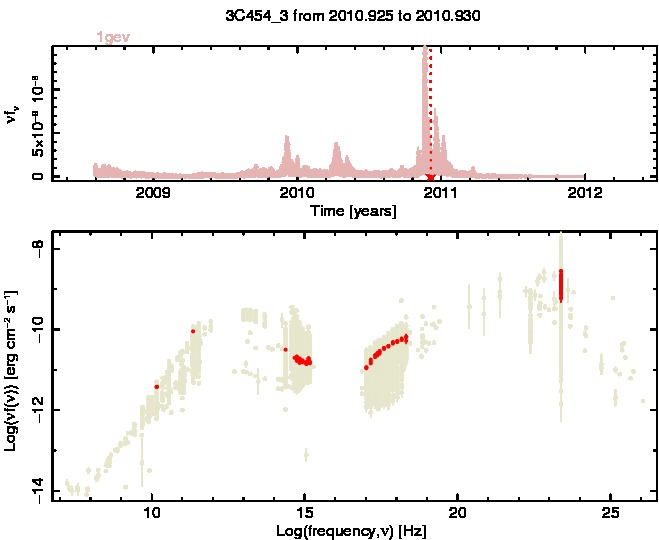}
\caption{A second frame from a SED movie of 3C454.3. The time interval corresponding to the red points in this case is 2010.825 - 2010.830, just after the maximum emission in the gamma-ray band.}
      \label{MovieFrame2}
\end{figure}

With this motivation, and in an attempt of implementing new methods of visualizing variability of the emission across the electromagnetic spectrum, I developed a prototype software tool to run 
in sequence SEDs corresponding to different time slices, that is to produce SED movies.

Figures \ref{MovieFrame1} and \ref{MovieFrame2} show two frames taken(stills) from the SED movie of 3C454.3 made with this prototype, corresponding to a period of approximately three days in late 2010 when the blazar underwent a very large gamma-ray flare. The data taken during the frame period is shown in red color, while all the remaining data is plotted in light gold color. The top panel shows the intensity of 3C454.3 in the gamma-ray band (1GeV) as a function of time and is used as a way to illustrate the passing of time. The bottom panel shows the full SED of the source using the same color coding for the data.  
This prototype will be further developed in the near future, likely as a collaborative effort among international institutions, and made openly accessible within the ASDC SED builder. 
A sample of a full SED movie of the blazar 3C454.3 is currently available on-line and can be seen at the  main page of the current version of the ASDC SED tool (V3.0).  

Usually blazar SED data is compared to theoretical models that are based on the current best understanding of the physical processes responsible for the emission of the multi-frequency radiation. When possible,  this is done using simultaneous data gathered through observational campaigns involving ground-based and satellite observatories.  However, as shown in Figs. \ref{MWlightcurves} and \ref{zdcf3c454.3} variability in different bands follow different dynamical timescales and may show significant time lags implying that fitting of simultaneous data may not be enough for a full comprehension of the physics at work in the source. 
 
New methods of analyzing data, that go beyond fitting simultaneous SEDs and fully take into account of the dynamical time-scales of the emission processes in different energy bands, require the development of new analysis tools. This is a challenge both for theoretician and scientific software developers to be addressed in the near future.
  
The advent of advanced visualization and fitting techniques such as SED movies and dynamical model fitting will likely happen soon, providing more diagnostic and interpretation power. 

\bigskip

\noindent {\bf Acknowledgments}

I acknowledge the use of archival data and software tools from the ASDC, a facility managed by the Italian Space Agency (ASI). Part of this work is based on archival data from the NASA/IPAC Extragalactic Database (NED).




\nocite{*}
\bibliographystyle{elsarticle-num}
\bibliography{giommi}






\end{document}